\documentclass[pra,reprint,superscriptaddress,showpacs,amsmath, floatfix,amssymb,aps]{revtex4-1}
\usepackage{graphicx}
\usepackage{subfigure}
\usepackage{dcolumn}
\usepackage{bm}
\usepackage{hyperref}
\usepackage{indentfirst}
\usepackage{braket}
\usepackage{amsmath}
\usepackage{color}
\usepackage{makecell}
\usepackage[ruled,vlined]{algorithm2e}
\usepackage{tabularx}
\usepackage{lipsum}
\hypersetup{colorlinks=true, citecolor=blue, urlcolor=blue, linkcolor=blue}
\begin{document}

\title{Beyond Conventional Pairing: Bosonic Quartic Superfluidity, Exotic XY Magnetism, and Phase Criticality }
\author{Chenrong Liu}
\affiliation{College of Mathematics and Physics, Wenzhou University, Zhejiang 325035, China}
\author{Zhi Lin}
\thanks{Corresponding author}
\email{zhilin18@ahu.edu.cn}
\affiliation{School of Physics and Optoelectronic Engineering, Anhui University, Hefei 230601, China}
\affiliation{State Key Laboratory of Surface Physics and Department of Physics, Fudan University, Shanghai 200433, China}

\begin{abstract} 
We report two unprecedented bosonic quartic superfluid (BQSF) phases in binary boson mixtures with synthetic pair-hopping (SPH) interaction realizable through Floquet engineering, transcending  the conventional bosonic superfluidity paradigm via high-order correlations.  Through extensive numerical simulations, we demonstrate: (i) The paired super-counter-fluid (PSCF) phase that exhibits bosonic `four-particle' correlations, directly mirroring fermionic 4e superconductivity; (ii) The symbiotic super-counter-fluid  which is an anomalous quantum phase featuring intrinsically intertwined, non-zero PSCF and super-counter-fluid orders.
Both phases further reveal unreported pseudo-spin-ordered XY ferromagnetic states with filling-dependent magnetization textures. Adjacent to  BQSF phases, we uncover both a previously unreported quantum quadruple critical point induced by interexchange asymmetry and SPH-driven state-dependent criticality—the latter exhibiting distinct behavior from conventional two-component bosonic systems.
We propose momentum-resolved noise correlation spectroscopy—building on established quantum gas microscopy techniques—as a direct probe for the characteristic signatures of these BQSF states. Our results establish a new paradigm in quantum matter, opening unprecedented avenues to explore (i) quartic quantum coherence, (ii) emergent collective phenomena, and (iii) extended XY magnetism—resolving a longstanding classification challenge while discovering fundamentally new phases beyond existing theoretical frameworks.

\end{abstract}
\maketitle
Ultracold atoms in optical lattices provide a versatile platform for simulating quantum many-body systems and exploring condensed matter phenomena~\cite{RevModPhys71.463,AdvPhys56.243,UltracoldAtomsinOpticalLattices,RepProgPhys82.104401}. Precise control over atomic interactions has enabled the observation of quantum phase transitions in
single-component bosonic systems, e.g., the superfluid–Mott insulator transition~\cite{Nature415} and the emergence of supersolid phases~\cite{Ss1,Ss2,Ss3,Ss4}. Intensive efforts currently focus on conventional two-component bosonic systems ~\cite{PhysRevLett90.100401,Demler,PhysRevLett92.050402,Nature425,PhysRevA77.011603,Nature588,PRL126,PRL128,SCF-detected, Nature425,PhysRevA77.011603,SCF-detected}, including the engineering of a  tunable Heisenberg  model and single-ion anisotropy terms  in  spin-1 models~\cite{Nature588,PRL126}, and the observation of novel phases such as the spin-Mott and super-counter-fluid (SCF) states~\cite{PRL128,SCF-detected}.
These systems are effectively described by the traditional  two-component Bose-Hubbard (TCBH) model, incorporating both intra- and interspecies on-site interactions and supporting rich phases such as the two-component Mott insulator (2MI), two-component superfluid (2SF), and `two-particle' pairing superfluid (SCF, and  paired superfluid (PSF))~\cite{PhysRevLett90.100401,PhysRevLett92.050402, PhysRevLett90.100401,Demler,PhysRevLett92.050402,PSF4,PSF5,PSF6,PSF7,PSF8,PSF9,PSF10}. Crucially, the SCF and PSF  represent novel interspecies pairing mechanisms absent in single-component bosonic systems. While most
studies assume interexchange symmetry between species, breaking this symmetry has been shown to introduce  additional complexity and enriches the phase diagram. A central question arises: Can such systems stabilize exotic bosonic interspecies pairing mechanisms beyond conventional paradigms—such as single-particle coherence  represented by Bose-Einstein condensation  (manifested as the 2SF state in TCBH) and two-particle (quadratic) coherence exemplified by paired superfluidity?

According to Schulz's taxonomy of anisotropic spin-S Heisenberg chains~\cite{Schulz},  XY1 ferromagnetic state is defined by non-vanishing correlations $\langle (S^{\dag})^n \rangle \neq 0$($S^{\dag}$ is spin-raising operator) for all integer $n \leq 2S$, whereas XY2 phase ($S > 1/2$) exhibits only $\langle (S^{\dagger})^{2S} \rangle \neq 0$ with suppressed lower-order fluctuations. Two decades ago, Kuklov et al. established the exact mapping between the XY1 phase and the SCF phase in bosonic systems~\cite{PhysRevLett90.100401}. This equivalence is physically realized through the Schwinger boson transformation: the finite order parameter $\langle ab^{\dagger} \rangle$ in the SCF phase directly implies transverse spin order $\langle S^{\dagger} \rangle \neq 0$, definitively confirming SCF as an XY1 state~\cite{PhysRevLett90.100401,Demler}.  This correspondence raises a fundamental question: Does the bosonic realm admit exotic XY2-type ordering beyond the XY1 paradigm? The potential realization of such XY2 order—and its conjectured connection to a new class of paired superfluidity—remains unexplored, forming the central motivation of this work.

In this work, we systematically investigate the ground-state phases and demixing transitions in the tunable TCBH+SPH framework. We reveal two distinct bosonic quartic superfluid (BQSF) phases induced by SPH interaction: (i) The novel paired super-counter-fluid (PSCF) phase manifests bosonic four-particle correlations with inter-species hole-particle duality (two bosons pairing with two holes), establishing the first bosonic analog of fermionic 4e superconductivity~\cite{4-e2,4-e3,4-e4}. (ii) The symbiotic super-counter-fluid (SSCF) phase represents a distinct quantum phase, where PSCF and SCF orders are inseparably intertwined. Both phases can unveil previously unreported types of XY ferromagnetic states, particularly pseudo-spin-ordered phases with unconventional magnetization textures. Significantly, we demonstrate that the interexchange asymmetry induces a quantum quadruple critical point, as well as the SPH interaction drives  state-dependent mixed-demixed criticality---a hallmark distinguishing our system from conventional two-component bosonic frameworks~\cite{PhysRevA76.013604,Iskin,JPhysSocJpn81.024001,PhysRevA89.057601,PhysRevA92.053610}.

\begin{figure*}[htbp]
\centering
\includegraphics[width=0.9\textwidth]{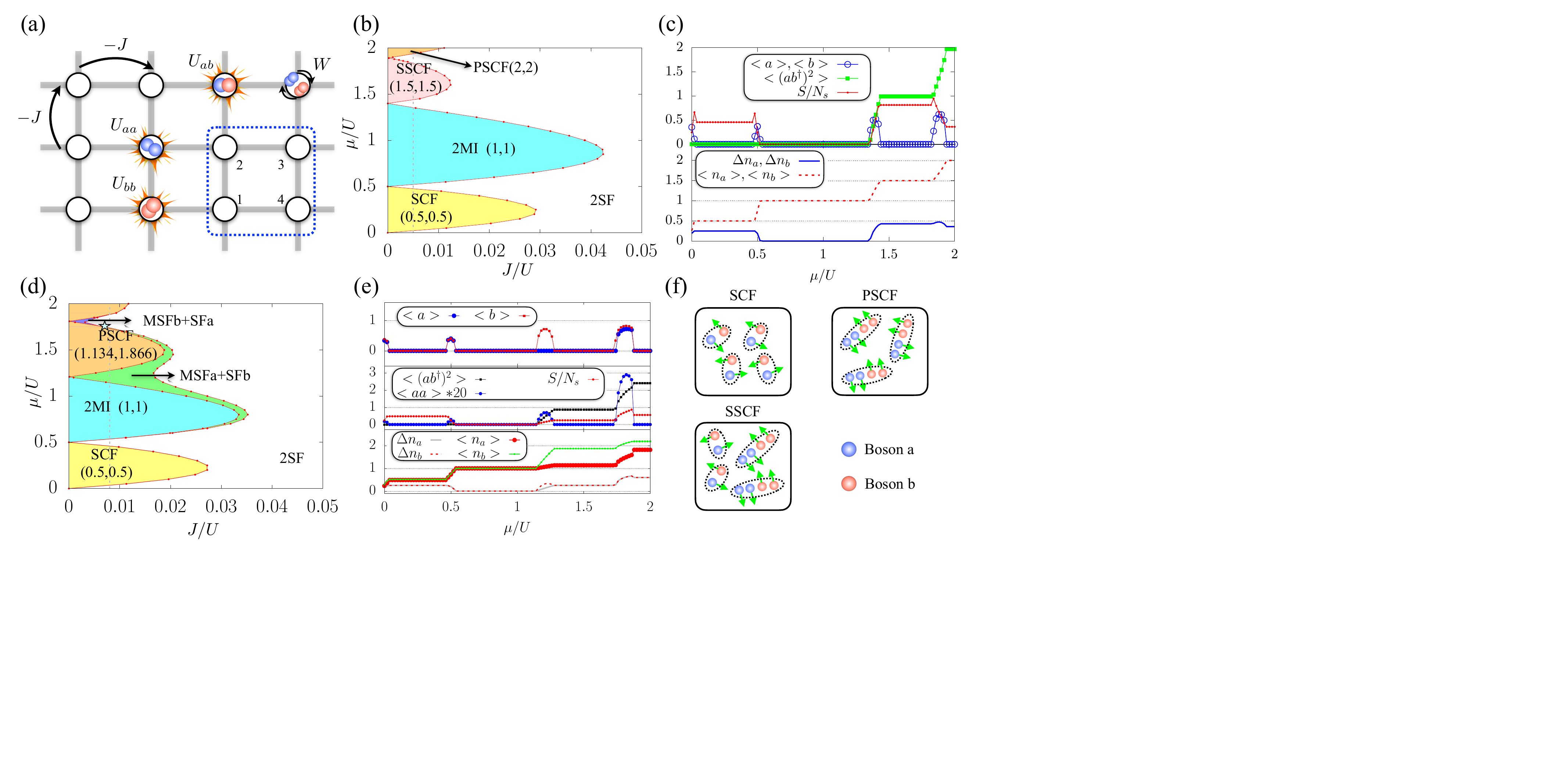}
\caption{\label{Fig3}(Color online) A sketch of TCBH+SPH  model and the corresponding phase diagrams (in units of $U_{aa}=U$).  (a) The TCBH+SPH model is illustrated in a square lattice, with the unit cell for CGMF calculations marked by a dashed blue box and internal sites labeled 1-4. Phase diagrams of chemical potential $\mu$ versus $J$ are shown for (b) $U_{bb}=U_{aa}=1$(interexchange symmetric case) and (d)$U_{bb}=0.8U_{aa}=0.8$ (interexchange asymmetric case), with fixed $U_{ab}=0.5$ and $W=-0.1$(These parameters is chosen based on comparison with previous studies ~\cite{PhysRevLett125.245301}). Subplots (c)/(e) display measurements along the black dashed lines in (b)/(d) at $J/U=0.005/0.008$. Brackets $(\langle n_a \rangle, \langle n_b \rangle)$ denote average particle numbers in phases, while a star in (d) marks the quantum quadruple critical point. (f) Schematic of SCF, PSCF, and SSCF phases: colors differentiate boson species, arrows indicate motion directions.}
\end{figure*}

\emph{Model---}Following Schulz’s definition, the minimum spin magnitude S required to realize the XY2 ferromagnetic phase is $S=1$, with the corresponding order parameter satisfying $\langle (S^{\dagger})^{2} \rangle \neq 0$ and $\langle S^{\dagger} \rangle = 0$.  A natural idea is that introducing an interaction of the form $ (S^{\dagger})^{2}+\text {H. c.}$ in a two-component bosonic system could realize this phase. Furthermore, applying the Schwinger boson transformation (such as mapping $ a_ib_i^\dagger $ to $S^{\dagger}_i$), the XY2 ferromagnetic phase corresponds directly to a quartic bosonic condensate characterized by the order parameter  $\langle (ab^\dag)^2 \rangle$. Therefore, in a two-component bosonic system, introducing this interaction not only facilitates the realization of the XY2 phase but also enables the discovery of novel superfluid states transcending the conventional bosonic superfluidity paradigm. We have recently proposed realizing this specific interaction ($ (S^{\dagger}_i)^{2} \mapsto (a_ib_i^\dagger)^2$) based on Floquet engineering~\cite{PhysRevLett125.245301,Bukov,Goldman,Eckardt_eff,Weitenberg}, termed the synthetic pair hopping (SPH) interaction~\cite{PhysRevLett125.245301}. Thus, we consider the following Hamiltonian of the TCBH+SPH model \cite{PhysRevLett125.245301}, 
\begin{eqnarray} \label{Ham}
H &=-J \displaystyle\sum_{\langle i,j\rangle}\left(a_{i}^{\dagger} a_{j}+b_{i}^{\dagger} b_{j}+\text {H. c.}\right)-\mu \sum_{i}\left(n_{i,a}+n_{i,b}\right) \nonumber\\ &+\displaystyle\frac{U_{aa}}{2} \sum_{i} n_{i,a}\left(n_{i,a}-1\right)+\frac{U_{bb}}{2} \sum_{i} n_{i,b}\left(n_{i,b}-1\right) \nonumber\\&+U_{ab}\displaystyle\sum_{i} n_{i,a} n_{i,b}+W \sum_{i}\left(a_{i}^{\dagger} b_{i} a_{i}^{\dagger} b_{i}+ \text {H. c.}\right),
 \end{eqnarray} 
 where $n_{i,a}$ ($n_{i,b}$) is the boson $a$($b$) number operator on site $i$, $J$  is the spin-independent hopping amplitude, $\mu$ is
spin-independent chemical potential, $U_{aa/bb}$($U_{ab}$) denotes the
strength of the on-site intraspecies  (interspecies) repulsive interactions  and $W$ is the strength of the SPH interaction, which can be realized via Floquet engineering \cite{PhysRevLett125.245301} and has also been suggested in a multi-band Bose system \cite{PhysRevB83.195106,PhysRevLett111.205302}. A sketch of this model in a square lattice is shown in Fig.~\ref{Fig3}(a).  
 



\emph{Phase diagram---}  Including the SPH interaction is expected to enrich the phase diagrams of the traditional TCBH~\cite{Yong-Shi-Wu,PhysRevLett92.050402,Iskin, PRA81.053608, arXiv2021}. 
By utilizing the cluster Gutzwiller mean-field method (CGMF)~\cite{PhysRevB86.144527,PhysRevA87.043619,JPhysB51.145302,PhysLettA129485}, we  firstly study the phase diagram in the interexchange symmetric case ($U_{bb}=U_{aa}$). As shown in Fig.~\ref{Fig3}(b) and (c), we reveal five distinct phases: the three established phases (2MI, SCF, and 2SF) along with two novel paired superfluid phases—PSCF and SSCF—extending beyond conventional bosonic pairing mechanisms. The pairing-pattern of SCF, PSCF, and SSCF is sketched in Fig.~\ref{Fig3}(f), and the corresponding behaviors of the order parameters are summarized in Table.~\ref{table1}.


In the PSCF phase, only the paired SCF order parameter $\langle (ab^\dag)^2 \rangle$ is non-zero, while the other order parameters such as $\langle a \rangle$, $\langle b \rangle$, $\langle ab^\dag \rangle_{J=0}$ and $\langle ab^\dag \rangle_{J\neq0}$ are all zero.  As we know, in an SCF state (the calculation details  via the CGMF  presented in SM~\cite{sm}), the net atomic superfluid current is zero,  but there exist equal superfluid currents of the components in opposite directions, each current composed of  single-atom transport \cite{PhysRevLett90.100401}. In stark contrast to  SCF phase, the PSCF phase exhibits zero net atomic superfluid current and vanishing counterpropagating currents driven by single-atom transport, yet sustains finite counterpropagating superfluid currents of equal magnitude governed by two-atom transport. Thus, the PSCF phase can be understood as a pairing between two particles from one component and two holes from the other component, which corresponds to a BQSF analogous to the fermionic charge-4e superconductor \cite{4-e2,4-e3,4-e4}.  Through Schwinger boson transformation, we establish the PSCF phase as a bosonic incarnation of the XY2 phase---a connection manifested by the emergent $\langle (S^{\dagger})^{2}\rangle \neq 0$ order.


\begin{table}[tb]
\caption{\label{table1}%
Identification of SCF, PSCF, and SSCF phases in terms of the values of order parameters $\langle \alpha \rangle$, $\langle ab^\dag \rangle$, $\langle (ab^\dag)^2 \rangle$, interspecies entanglement entropy $S$, and particle number fluctuations $\Delta n_{a/b}$. Reminder that all these observables are calculated based on the CGMF ground states and have been averaged by the size of the cluster.
}
\begin{ruledtabular}
\begin{tabular}{lllllll}
Phases & $\langle \alpha \rangle$  & $\langle ab^\dag \rangle_{J=0}$ &$\langle ab^\dag \rangle_{J\neq0}$& $\langle (ab^\dag)^2 \rangle$ & $S$ & $\Delta n_{a/b}$\\
 \hline
SCF & 0 & $\neq 0$ & 0& 0 & $\neq 0$ & $\neq 0$\\
PSCF & 0 & $0$ & 0 & $\neq 0$ & $\neq 0$ & $\neq 0$\\
SSCF & 0 & $\neq 0$ & 0 & $\neq 0$ & $\neq 0$ & $\neq 0$\\
\end{tabular}
\end{ruledtabular}
\end{table}

Interestingly, unlike the SCF or 2SF phases, which request non-integer averaged single-component particle numbers,  the PSCF($\langle n_{a}\rangle,\langle n_{b}\rangle $) phase can have integer filling for each component $\langle n_{\alpha} \rangle$ ($\alpha\!=\!a,b$), such as  $\langle n_{a} \rangle=\langle n_{b} \rangle=2$  shown in Fig.~\ref{Fig3} (b). This integer value results from two constraints: a) The conserved total particle number in the PSCF lobe is even (e.g. the filling factors \!$f=\langle n_{ a} \rangle \!+ \!\langle n_{ b} \rangle\!=\!4$);  b) interexchange symmetry is preserved, i.e. requiring $\langle n_{a} \rangle  =\langle n_{b} \rangle $. In addition, the PSCF lobe exhibits constant $\langle n_{\alpha} \rangle$ with increasing $\mu$, mimicking the characteristic behavior of a 2MI state. However, as shown in Figs.~\ref{Fig3}(c) and (e), we discover that the PSCF phase combines finite interspecies entanglement entropy ($S\!>\!0$) and nonzero compressibility ($\Delta n_{\alpha}\!\neq \!0$), in contrast to the 2MI phase ($S=\!0$, $\Delta n_{\alpha}\!=\!0$).    
We explore the effects of breaking interexchange symmetry by considering $U_{bb}\!=0.8\!<\!U_{aa}\!=\!1$ as an example.
As depicted in Figs.~\ref{Fig3}(d-e), the PSCF phase persists but exhibits non-integer fillings $\langle n_{\alpha} \rangle$ for each component, contrasting the symmetric case ($U_{bb}=U_{aa}$) with integer $\langle n_{\alpha} \rangle$.  The reason is that although  the filling factor $f$ is a conserved integer, $\langle n_{a} \rangle$ and  $\langle n_{b}\rangle$ must be unequal in the asymmetric case ($U_{bb}\neq U_{aa}$), and their values depend on the imbalance of $U_{aa}$ and $U_{bb}$. Thus, the PSCF phase in this scenario corresponds to non-integer Mott insulator state~\cite{PhysRevLett125.245301}. 

On the other hand, the SSCF phase is characterized by non-zero values for both $\langle ab^\dag \rangle$ and $\langle (ab^\dag)^2 \rangle$, while all the other order parameters remain zero. This indicates that the SSCF phase combines the features of both the SCF and PSCF states, representing a novel paired superfluid phase that has not been previously studied. At $W=0$, the SSCF(1.5,1.5) phase observed in Fig.~\ref{Fig3}(b) undergoes transition to the conventional SCF (1.5,1.5) state characterized in Supplementary  Fig.~S2~\cite{sm}. Thus, the existence of the SSCF phase is attributed to the $W$ term. 

Moreover, the interspecies interaction $U_{ab}$ can significantly influence the phase diagram. 
At the atomic limit ($J=0$), we theoretically analyze the second lobe's ground state and identify a critical value of $U_{ab}$, given by $U_{ab}^c=U_{bb}+2Wk$, where $k=(U_{aa}-U_{bb})/4W \pm 1/2 \sqrt{(U_{bb}-U_{aa})^2/4W^2+4}$ is a parameter (the calculation details in SM~\cite{sm}). For $U_{ab}<U_{ab}^c$, the second lobe corresponds to the 2MI phase, whereas for $U_{ab}>U_{ab}^c$, it transitions to a PSCF state. In particular, at $U_{ab}=U_{ab}^c$, the lobe can exhibit an SSCF phase,  regardless of  whether the system is interexchange symmetric or asymmetric.

 \emph{Quantum quadruple criticality---}Notably, we find a quantum quadruple critical point (marked by stars)  in Fig.~\ref{Fig3}(d). This novel quantum quadruple critical point represents a phase transition among the 2SF, PSCF,  molecular superfluid of  species a + superfluid of species b (MSFa+SFb)~\cite{PhysRevLett125.245301}, and MSFb+SFa.  To the best of our knowledge, it is the first observation of a quantum quadruple critical point revealed in a TCBH+SPH model. Additionally,  in the regime of small chemical potential $\mu$, several quantum quadruple critical points  emerge for weak inter-species interactions $U_{ab}=0.1$ (see Fig.~S5 (h) in SM.~\cite{sm}).  This originates from the reduction of phase space for SSCF and SCF states as $U_{ab}$ decreases when $U_{aa}=U_{bb}$. Breaking the interexchange symmetry ($U_{aa}\neq U_{bb}$) transforms the SSCF phase into a PSCF phase, enabling multiple PSCF states at small $\mu$. Each PSCF lobe then develops a quantum quadruple critical point near its tip, resulting in multiple such critical points in the system. 

\begin{figure}[htbp]
\centering
\includegraphics[width=0.45\textwidth]{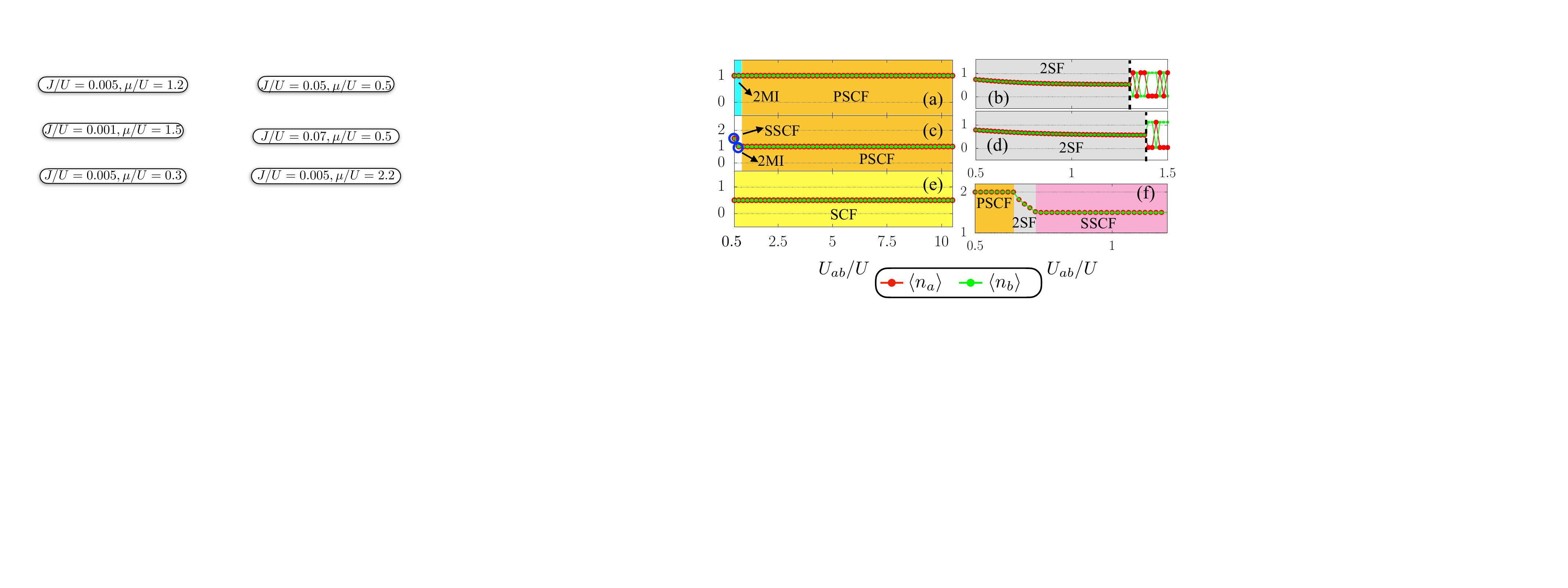}
\caption{\label{Fig7}(Color online) Demixing effect at $W/U=-0.1$ with the interexchange symmetry is preserved. Here, we start tuning the $U_{ab}/U$ from 0.5 which is set in Fig.~\ref{Fig3}. Other model parameters $(J/U,\mu/U)$ are fixed at (a)$(0.005,1.2)$, (b)$(0.05,0.5)$, (c)$(0.001,1.5)$, (d)$(0.07,0.5)$, (e)$(0.005,0.3)$, and (f)$(0.005,2.2)$. The black dashed vertical lines in (b) and (d) mark the mixing-demixing phase transition points. }
\end{figure}
\emph{Emergent XY Ferromagnetism---} The PSCF and SSCF phases can be reinterpreted as novel XY ferromagnet phases within the pseudospin framework. Following Schulz’s classification \cite{Schulz}, the XY ferromagnet phase in spin-S systems splits into two types: XY1 and XY2. 
By mapping filling factors $f$  to effective spins $S=f/2$ \cite{PhysRevLett90.100401,PhysRevLett.91.090402}, distinct PSCF phases correspond to specific XY ferromagnet types. For instance, the  PSCF phase with $f=2$ ( see Figs.~S5(c) and (f) in SM~\cite{sm}) maps to a spin-1 XY2 phase ($\langle (S^{\dag})^2\rangle \neq 0$ and $\langle S^{\dag}\rangle =0$), while other PSCF phases, e.g., PSCF(2,2) and PSCF(1.134,1.866) in Figs~\ref{Fig3} (b) and (d), represent a new XY-type phase with  $\langle (S^{\dag})^2\rangle \neq 0$ but vanishing lower/higher moments  $\langle S^{\dag}\rangle=\langle (S^{\dag})^3 \rangle =\langle (S^{\dag})^4 \rangle=0$, deviating from standard XY1/XY2 criteria \cite{Schulz}. Similarly, the  SSCF phase  with $f=2$ (see Figs.~S5(b) and (e) in SM~\cite{sm}) corresponds to a spin-1 XY1 phase, whereas the  SSCF phase with $f=3$ (see Fig.~\ref{Fig3}(b))  exhibits a hybrid order($\langle S^{\dag}\rangle \neq 0$, $\langle (S^{\dag})^2\rangle \neq 0$ but $\langle (S^{\dag})^3\rangle = 0$), further defining a distinct XY ferromagnet beyond conventional classifications.

\emph{Phase-dependent demixing---}Strong $U_{ab}$ triggers demixing~\cite{PhysRevA76.013604, Iskin, JPhysSocJpn81.024001, PhysRevA89.057601, PhysRevA92.053610}, causing phase separation (or collapse) and instability. We analyze the evolution of these phases under large $U_{ab}$ in the demixed regime, considering the interexchange-symmetric case ($U_{aa}=U_{bb}=1$). We can expect that in all mixed phase, the equality $\langle n_a \rangle=\langle n_b \rangle$ holds. But in a complete demixing phase, species separation prevents onsite coexistence, leading to $\langle n_a \rangle \neq \langle n_b \rangle$ (serving as evidence of demixing) and spontaneous interexchange symmetry breaking. 

For $W\!\!=\!\!0$, two species mix stably under the condition  $U_{ab}^2 \!<\!U_{aa}U_{bb}$ with repulsive interactions $U_{\alpha \alpha}\!>\!\!0$  \cite{Iskin}. Consequently, in the interexchange symmetric case, the conventional two-component bosonic system undergoes demixing transition at a phase-independent critical interspecies interaction $U_{ab}^D=1$ \cite{PhysRevA76.013604, JPhysSocJpn81.024001, PhysRevA89.057601, PhysRevA92.053610} (See Fig.~S4~\cite{sm}). Contrasting with the  $W\!\!=\!\!0$ limit, finite SPH interaction ($W\neq 0$) can modifies the mixed-demixed transition and it lowers energy via interspecies pair hopping process, which stabiliz the mixed state. Since hopping strength increases with particle density,  the demixing critical point  $U_{ab}^D$ becomes phase-dependent when $W$ is present. To verify this, we choose $W=-0.1$ and initialize  $(J,\mu)$ to locate the 2MI, 2SF, SCF, PSCF, and SSCF phases  at $U_{ab}=0.5$, respectively, and track the evolution of demixing in these states with increasing  $U_{ab}$.  As shown in Fig.~\ref{Fig7}, our analysis reveals distinct phase evolution pathways: (i) Conventional phase transitions occur in most phases (the 2MI-to-PSCF transition  in Fig.~\ref{Fig7}.(a), the sequential SSCF-to-2MI-to-PSCF evolution  in Fig.~\ref{Fig7}.(b), and the PSCF-to-2SF-to-SSCF progression  in Fig.~\ref{Fig7}.(f)), while the 2SF phase exhibits exclusive demixing; (ii) The SCF lobe's $\Delta \mu  (J=0)$ increases linearly with $U_{ab}$ below the critical value ($\Delta \mu = U_{ab}$ for $ U_{ab}<U_{ab}^c $), then plateaus at $\Delta \mu = U_{ab}^c$ beyond this threshold. Consequently, for parameters ($J=0.05,\mu=0.3$)  the system remains stably within the SCF phase throughout $U_{ab}$ variation without undergoing phase transitions (Fig.~\ref{Fig7}.(e)).  Through mean-field analysis in the deep 2SF regime, we derive the critical interaction strength $U_{ab}^D = U - 2W + 4J/nL$, where $n$ is the single-component superfluid density at the critical point and $L$ is the linear system size (for square lattices, the lattice sites $N$ satisfy $N=L^2$). Our analytical prediction ($U_{ab}^D = 1.39$ for Fig.~\ref{Fig7}.(b); $1.44$ for Fig.~\ref{Fig7}.(d)) agrees remarkably with our numerical results ($1.32$ and $1.40$ respectively). The observed dependence of $U_{ab}^D$ on $J$ and $n$ (See Fig.~\ref{Fig7}.(b) and (d)) originates from finite-size effects,  and it will disappear in thermodynamic limit ($L\rightarrow \infty$). These results establish that the $W$-term induces state-dependent mixing-demixing transitions—a new mechanism beyond conventional phase separation paradigms.

\begin{figure}[htbp]
\centering
\includegraphics[width=0.45\textwidth]{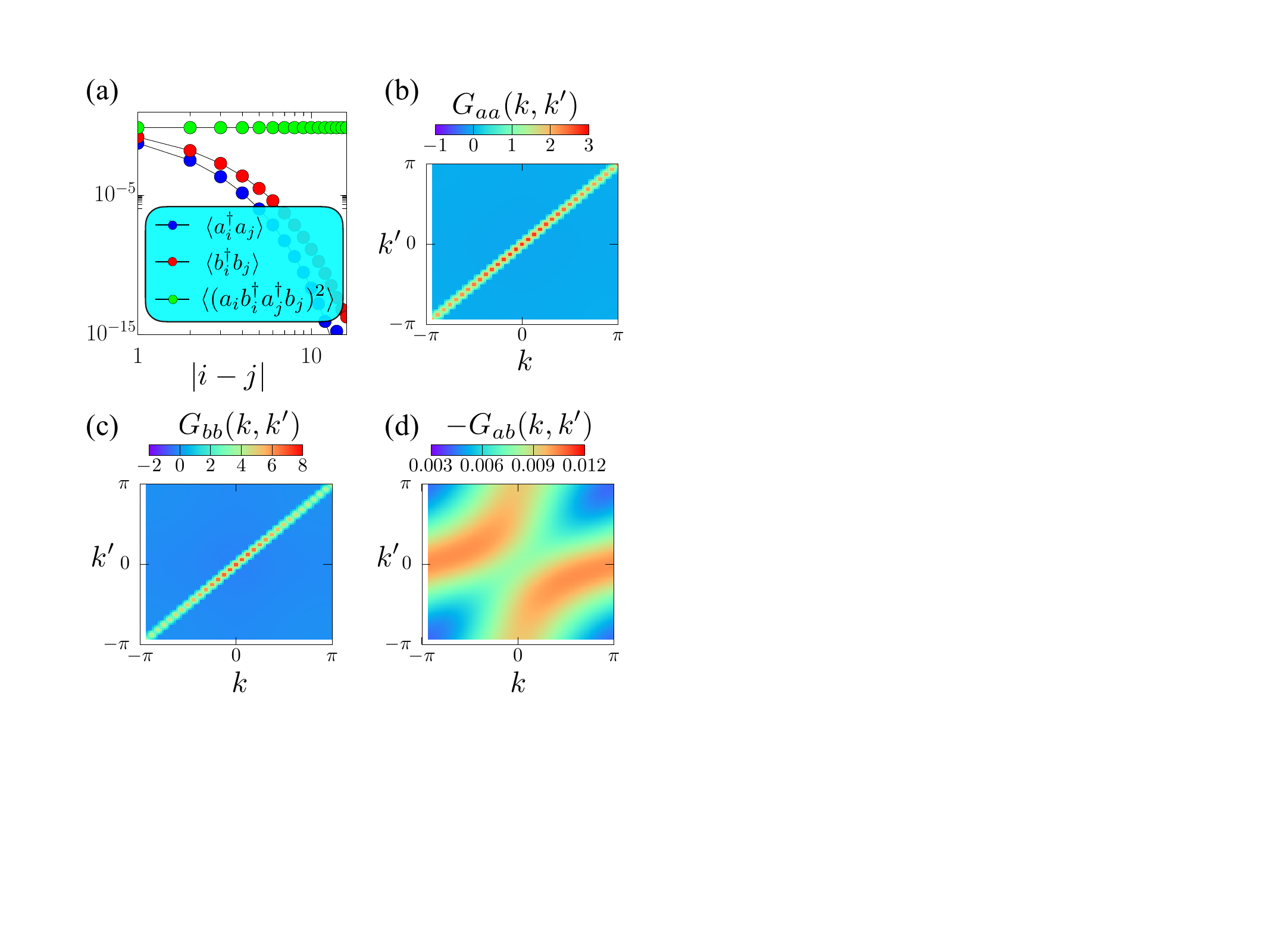}
\caption{\label{Fig9}(Color online) DMRG results for (a) correlation functions and (b)-(d) noise correlations in the PSCF phase of a 1D TCBH+SPH chain. The chain length is fixed at  $L=32$, with global parameters $J/U_{aa}=0.01$, $U_{bb}/U_{aa}=0.8$, $U_{ab}/U_{aa}=0.5$ and $W/U_{aa}=-0.1$. Within the canonical ensemble, the total particle filling factor $(N_a+N_b)/L=3$ is enforced. Notably, $N_a$ and $N_b$ fluctuate due to the $W$ term, the total particle number $N=N_a+N_b$ remains conserved. }
\end{figure}
\emph{ Experimental detection---}The SCF phase—a novel `two-particle' pairing superfluid originally proposed two decades ago and equivalent to the XY1 ferromagnetic state—has recently been confirmed experimentally via interspecies antipair correlations~\cite{SCF-detected}. Specifically, this phase manifests strong antipair correlations at equal momenta ($k=k^\prime$)  in the interspecies noise correlators, defined as $G_{a,b}=\langle n_{a,k}n_{b,k^{\prime}}\rangle -\langle n_{a,k}\rangle \langle n_{b,k^{\prime}}\rangle$~\cite{SCF-detected,PhysRevA.81.063602}. These pronounced equal-momentum correlations serve as a diagnostic signature of the SCF phase. This experimental approach provides an excellent platform for probing the two distinct BQSF phases predicted theoretically in this work.  We propose feasible experimental protocols to identify the two distinct BQSF phases.

The PSCF phase can be identified via distinct noise correlation signatures and Feshbach resonance techniques. We find that the PSCF phase exhibits Mott-like intraspecies correlations in both $G_{a,a}$ (see Fig.~\ref{Fig9}(b)) and $G_{b,b}$ (see Fig.~\ref{Fig9}(c)) with peaks along  $k=k^\prime$ direction, while $G_{a,b}$ displays arc-shaped correlations along $(-\pi,0)\rightarrow (0,\pi)$ and $(0,-\pi)\rightarrow (\pi,0)$ (see Fig.~\ref{Fig9}(d)), contrasting the linear antipairing in SCF. Crucially, pairing features emerge when probing Feshbach molecules: applying a magnetic field to form  $aa$ and  $bb$ molecules, a rapid Feshbach ramp reveals antipairing in the molecular correlator $F_{a,b}=\langle n^{M}_{a,k}n^{M}_{b,k^{\prime}}\rangle -\langle n^{M}_{a,k}\rangle \langle n^{M}_{b,k^{\prime}}\rangle$, where peaks along $k=k^\prime$ signal $a_ja_jb_j^\dagger b_j^\dagger$  pairing, confirming PSCF. Here, the number operator for $\alpha \alpha $  Feshbach molecule in momentum space is $n^{M}_{\alpha,k}=M_{\alpha,k}^{\dagger}M_{\alpha,k}$, with  real-space annihilation operator $M_{\alpha,j}=\alpha_j\alpha_j$ ($\alpha\!=\!a,b$). In the PSCF phase, the noise correlation $F_{a,b}$ contains $f_{A}(k,k^{\prime})$, the Fourier transform of the molecular correlator $R_A^{M}(j_1,j_2)=\langle M_{a,j_1}^{\dagger} M_{a,j_2}M_{b,j_2}^{\dagger}M_{b,j_1}\rangle$, given by  $f_{A}(k,k^{\prime})=\sum_{j_1,j_2}R^{M}_{A}(j_1,j_2)e^{i(k-k^\prime)(j_1-j_2)}$. Peaks along $k=k^\prime$ signify local $M_{a,j}M_{b,j}^{\dagger}$ pairing, corresponding to $a_ja_jb_j^\dagger b_j^\dagger$ pair formation. These noise correlation features also consistent with the decay behaviors of correlation functions in Fig.~\ref{Fig9}(a).

Experimentally, a  $^{87}$Rb mixture in states  $|f=1, m_f=1\rangle$ and $|f=1, m_f=0\rangle$ under $\sim 410G$ magnetic fields can access Feshbach resonances for creating molecules in channels $(f_{1}=1, f_{2}=1) \nu=-2$ (where $\nu$ is a vibrational quantum number) and $(f_{1}=1, f_{2}=2) \nu=-4$~\cite{PhysRevLett.89.283202}. This setup, combined with SPH interactions, provides a viable platform for probing antipairing order in PSCF. Extending this approach to the SSCF phase—which hosts coexisting PSCF and SCF orders—requires integrating both molecular antipairing ($F_{a,b}$) and interspecies antipairing ($G_{a,b}$) measurements. 
\emph{Conclusions---}Using the CGMF method, we systematically study the ground-state phase diagram of the TCBH model with SPH interactions, identifying two novel BQSF phases—PSCF and SSCF—that exhibit emergent pseudospin XY ferromagnetic orders. Importantly, the strength of the interspecies interaction $U_{ab}$  restructures the ground-state phase diagram, shifting phase boundaries  and reconfiguring the domains of interspecies paired superfluids (PSCF and SSCF). Notably, a quantum quadruple critical point can emerge in the asymmetric case. Moreover, the number of such critical points can be controlled by tuning the value of $U_{ab}$, and this represents the first reported instance of a quantum quadruple critical point within the TCBH+SPH framework.


Notably, phase demixing exhibits phase-dependent behavior: strong $U_{ab}$ induces demixing exclusively in the 2SF phase, while other phases remain mixed. This contrasts with the traditional TCBH model, where all phases demix simultaneously at a single critical point. We propose universal detection protocols for exotic superfluid phases (MSFa, PSCF, SSCF) via momentum-space noise correlations ($G_{a,b}$) and molecular antipairing signatures ($F_{a,b}$), enabling unambiguous identification of pairing/antipairing features.

\emph{ Acknowledgements---} The authors thank Y. Chen, J. Lou, Botao Wang and Mingfeng Wang for fruitful discussions.  This research is supported by the National Natural Science Foundation of China (NSFC) under Grant No.12004005, the Scientific Research Fund for Distinguished Young Scholars of the Education Department of Anhui Province No.2022AH020008, the Natural Science Foundation of Anhui Province under Grant No. 2008085QA26,  the Scientific Research Fund of Zhejiang Provincial Education Department under Grant No. Y202248878, the Ph.D. research Startup Foundation of Wenzhou University under Grant No. KZ214001P05, and the open project of the state key laboratory of surface physics in Fudan University (Grant No. KF2021$\_$08 and Grant No. KF2022$\_$06).

\bibliography{Manuscript}

\end{document}